\documentclass[conference]{IEEEtran}
\IEEEoverridecommandlockouts

\usepackage{cite}
\usepackage{amsmath,amssymb,amsfonts}
\usepackage{graphicx}
\usepackage{textcomp}
\usepackage{xcolor}
\usepackage{booktabs}
\usepackage[hidelinks]{hyperref}
\usepackage{algorithm}
\usepackage{algpseudocode}

\usepackage{graphicx}
\usepackage{subcaption}

\usepackage{newtxtext,newtxmath}

\newtheorem{lemma!}{Lemma}
\newtheorem{corollary!}{Corollary}

\def\BibTeX{{\rm B\kern-.05em{\sc i\kern-.025em b}\kern-.08em
    T\kern-.1667em\lower.7ex\hbox{E}\kern-.125emX}}
\begin{document}

\title{Pipelining Kruskal's: A Neuromorphic Approach for Minimum Spanning Tree}



\author{\IEEEauthorblockN{Yee Hin Chong}
\IEEEauthorblockA{\textit{Department of Computer Science \& Technology} \\
\textit{Tsinghua University}\\
Beijing, China \\
yuxuanzh23@mails.tsinghua.edu.cn}
\and
\IEEEauthorblockN{Peng Qu}
\IEEEauthorblockA{\textit{Department of Computer Science \& Technology} \\
\textit{Tsinghua University}\\
Beijing, China \\
qp2018@mail.tsinghua.edu.cn}
\and
\IEEEauthorblockN{Yuchen Li}
\IEEEauthorblockA{\textit{Department of Computer Science \& Technology} \\
\textit{Tsinghua University}\\
Beijing, China \\
liyuchen24@mails.tsinghua.edu.cn}
\and
\IEEEauthorblockN{Youhui Zhang}
\IEEEauthorblockA{\textit{Department of Computer Science \& Technology} \\
\textit{Tsinghua University}\\
Beijing, China \\
zyh02@tsinghua.edu.cn}
}

\maketitle

\begin{abstract}

Neuromorphic computing, characterized by its event-driven computation and massive parallelism, is particularly effective for handling data-intensive tasks in low-power environments, such as computing the minimum spanning tree (MST) for large-scale graphs. The introduction of dynamic synaptic modifications provides new design opportunities for neuromorphic algorithms. Building on this foundation, we propose an SNN-based union-sort routine and a pipelined version of Kruskal’s algorithm for MST computation. The event-driven nature of our method allows for the concurrent execution of two completely decoupled stages: neuromorphic sorting and union-find. Our approach demonstrates superior performance compared to state-of-the-art Prim 's-based methods on large-scale graphs from the DIMACS10 dataset, achieving speedups by 269.67x to 1283.80x, with a median speedup of 540.76x. We further evaluate the pipelined implementation against two serial variants of Kruskal’s algorithm, which rely on neuromorphic sorting and neuromorphic radix sort, showing significant performance advantages in most scenarios.

\end{abstract}


\section{Introduction}

Neuromorphic computing leverages massive parallelism and event-driven computation, making it an effective paradigm for parallel acceleration, particularly in machine learning and graph learning on non-Von Neumann architectures \cite{survey,Fang2023}. The simplest design of a neuromorphic algorithm involves embedding the computational kernel directly into a static, non-modifiable spiking neural network (SNN), which is then deployed on neuromorphic hardware \cite{Hamilton2019,Hamilton2020}. This method sacrifices flexibility in exchange for significant gains in energy-efficient execution \cite{Kwisthout2020,Aimone2021}.

The introduction of various learning mechanisms, such as synaptic plasticity \cite{Shouval2010,Saponati2023,Lu2024}, has fostered the development of self-adaptive neuromorphic primitives. These strategies not only enhance biological plausibility but also significantly improve computational power.

A noteworthy advancement in this field is structural plasticity, which involves the dynamic formation, modification, and elimination of synaptic connections, or in short, synaptic rewiring\cite{Poirazi2001}. This mechanism creates new opportunities for designing neuromorphic algorithms. With structural plasticity, the local connectivity of an SNN can be dynamically adjusted based on the algorithm's needs \cite{vanOoyen2017}, similar to pointer manipulation in traditional computing. Such flexibility enables the creation of novel neuromorphic operators and algorithms \cite{Diazpier2016,Hussain2016,Janzakova2023,Zyarah2024}.

However, existing analyses of computational complexity in neuromorphic algorithms \cite{complexity,Kwisthout2020,Aimone2021} do not account for the overhead introduced by such learning rules. To address this, we propose a revision of the neuromorphic time complexity proposed in \cite{complexity}, extending conventional analysis to include the costs associated with structural modifications. This updated framework offers a more accurate characterization of algorithmic performance, which will be further discussed in this section.

Building on the concept of structural plasticity, we demonstrate how this learning rule can be used to design more efficient neuromorphic algorithms. We use the minimum spanning tree (MST) construction, specifically the kernel of single-linkage clustering \cite{Murtagh2012,Gagolewski2024} in machine learning, as a case study. By leveraging Kruskal's algorithm\cite{Kruskal1956}, we develop a union-find routine based on SNN primitives, and evaluate its performance compared to state-of-the-art approaches that use Prim’s algorithm \cite{Kay2020,Janssen2024} in the context of neuromorphic sorting and neuromorphic radix sorting.

We explore how spike-driven computation in neuromorphic systems facilitates parallelization opportunities for \textit{pipelining Kruskal’s algorithm}. This approach helps overcome the performance bottlenecks that arise from the sequential execution of sorting and union-find operations, which are typically independent. We take the initial steps in designing a pipelined version of Kruskal's algorithm that utilizes neuromorphic primitives along with the principles of structural plasticity. 

Extensive experiments on the DIMACS10 dataset \cite{SparseSuite} show that the pipelined Kruskal's outperforms the state-of-the-art Prim 's-based methods with a median speedup of 540.76x. Moreover, the results reveal that, in most cases, pipelining results in significant performance improvements over the sequential approaches. We also examine scenarios when pipelining may face bottlenecks, potentially leading to a decline in performance compared to sequential execution, and propose methods to identify such cases, supported by concrete examples.

The remainder of the paper is organized as follows: Section \ref{subsec:complex} discusses the revision of the complexity framework based on structural plasticity, and Section \ref{subsec:sort} presents implementations for both neuromorphic sorting and neuromorphic radix sorting. In Section \ref{sec:design}, we introduce the design of the union-find routine and the pipelined Kruskal's algorithm, evaluating their operational costs with a summary in Table \ref{tab:cost}. Section \ref{sec:results} presents experimental results on the DIMACS10 dataset, highlighting the conditions under which the sequential approach outperforms the pipelined version. Finally, Section \ref{sec:discuss} examines the feasibility of implementing these algorithms on neuromorphic hardware.

\section{Background}

\subsection{Neuromorphic Computing Complexity}
\label{subsec:complex}

The evaluation process on neuromorphic algorithms begins with neuromorphic graph primitives \cite{Hamilton2019, Hamilton2020}, which incorporate a graph of nodes and edges into an SNN using leaky integrate-and-fire (LIF) neurons and static synapses \cite{NeuroDyn}. This configuration provides a Turing-complete mathematical model for assessing the performance of a neuromorphic algorithm \cite{Date2022}. Building on earlier research, \cite{complexity} introduces a theoretical framework to define the computational complexity of these neuromorphic algorithms.

In this model, neurons accumulate signals from incoming synapses until they reach a predefined threshold, denoted $\upsilon_i$. Once this threshold is reached, the neuron emits a spike, transmits signals through outgoing synapses, and resets its state to zero. Each neuron has a leak factor, $\lambda_i$, which indicates how quickly it returns to zero if it does not spike. Both $\upsilon_i$ and $\lambda_i$ are whole numbers. A synapse processes the incoming signals from the pre-synaptic neuron $\mathrm{i}$ by multiplying them by its weight $\omega_{i,j}$, applies a delay $\delta_{i,j}$, and then delivers the signals to the post-synaptic neuron $\mathrm{j}$. The weights are integers, whereas the delays are non-negative numbers. Despite several variations, the general computing paradigm of a LIF neuron could be summarized as $v_i =  v_i / \lambda_i + w_{ij}$, where $v_i$ represents the signal accumulated in the membrane potential \cite{Fang2023}, and the input does not decay.

\begin{figure}
    \centering
    \includegraphics[trim=95 590 175 100, clip,width=0.8\columnwidth]{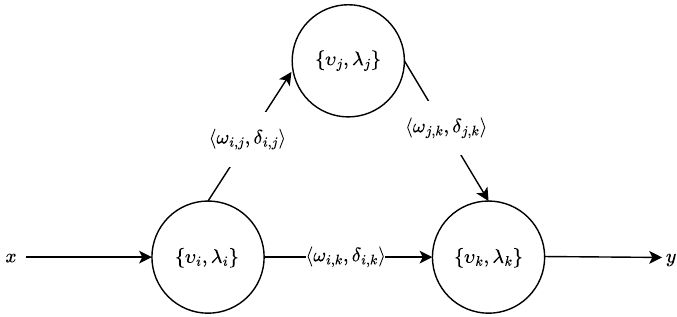}
    \caption{Symbolic notation for a typical SNN}
    \label{fig:symbolic}
\end{figure}

Figure \ref{fig:symbolic} illustrates the symbolic notation for an SNN. The circles labeled $\left\{\upsilon_i, \lambda_i\right\}$ represent neurons, while the arrows $\left<\omega_{i,j}, \delta_{i,j}\right>$ represent the synapses that connect them. Each pair of neurons is linked by a single synapse, as dictated by the framework specifications. The complexity reflects the resources required to configure and execute the SNN, consistent with the general definition provided by \cite{Arora2009}.

\begin{figure}
    \centering
    \includegraphics[trim=127 555 250 108, clip, width=0.8\columnwidth]{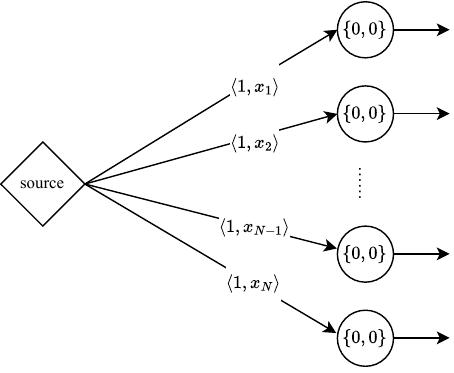}
    \caption{SNN of \texttt{NeuroSort}}
    \label{fig:sorting}
\end{figure}

The \textbf{time complexity}, denoted as $T(n)$, combines the \textbf{setup time} and the \textbf{running time} using the conventional big-O notation. The setup time refers to the duration required to configure neurons and synapses sequentially, which is proportional to the size of the SNN. In contrast, the run time is the period from when the inputs ($x$'s) are fed into the network until valid outputs ($y$'s) are produced. \textbf{Space complexity}, represented as $S(n)$, depends on the number of neurons and synapses within the SNN. Furthermore, \cite{Kwisthout2020} proposes the \textbf{energy complexity}, $E(n)$, to measure the total spike count used in the algorithm. In practice, $E(n) \leq T(n)\cdot S(n)$ since any neuron can fire at most once per time step.

Building on the data movement analysis proposed in \cite{Aimone2021}, to account for the actual structural plasticity during execution, we allow suspension of the neuromorphic activity of a subset of neurons and synapses (assumed to be $k$ in number) during runtime to modify connections. The directionality of these synapses is then modified, similar to pointer adjustments done in $O(1)$ time. The cost of these modifications still adheres to the setup time constraint, i.e., $O(k)$. 

\subsection{Sorting using an SNN}
\label{subsec:sort}

Given an input array of whole numbers, denoted as
\[
    \begin{aligned}
        \boldsymbol{x} = \left\{x_1,x_2,\cdots,x_{N-1},x_N\right\}
    \end{aligned}
\]

with $N$ elements, we aim to sort these numbers in ascending order using a spiking mechanism \cite{complexity}. An SNN is configured with synaptic delays defined by the elements of $\boldsymbol{x}$, as illustrated in Figure \ref{fig:sorting}. When the spike source is activated, the neurons, with $\upsilon_i = 0$, will fire according to their respective delays, resulting in the inputs being presented in a sorted sequence. Notice that both $w_{ij}$ and $\lambda_i$ are idle, and setting them to any non-negative value will not affect the computation. The pseudocode for this sorting kernel is provided in Algorithm \ref{algo:sort-naive}.

\begin{algorithm} 
\caption{Neuromorphic Sort}
\label{algo:sort-naive}
\begin{algorithmic}
    \Function{NeuroSort}{$arrIn$}
        \State $\mathrm{arrOut} \gets \mathopen{}\left[  \mathclose{}\right]$
        \State $t \gets 0$
        \While{$\mathrm{len} \mathopen{}\left( \mathrm{arrOut} \mathclose{}\right) < \mathrm{len} \mathopen{}\left( \mathrm{arrIn} \mathclose{}\right)$}
            \For{$v \in \mathrm{arrIn}$} \Comment{neuromorphic parallelism}
                \If{$v = t$}
                    \State $\mathrm{arrOut}.\mathrm{append} \mathopen{}\left( v \mathclose{}\right)$
                \EndIf
            \EndFor
            \State $t \gets t + 1$ \Comment{go to next timestep}
        \EndWhile
        \State \Return $\mathrm{arrOut}$
    \EndFunction
\end{algorithmic}
\end{algorithm}

The space complexity and the setup time for the algorithm are both $O(N)$, while the running time is $O(\max \boldsymbol{x})$, which refers to the largest element in $\boldsymbol{x}$. It's important to note that the run time is also bounded by $O\left(2^b\right)$, where $b = \log_2\left(\max \boldsymbol{x}\right)$ represents the number of bits needed to store this largest element. This indicates that the algorithm operates in pseudo-polynomial time \cite{GareyJ79}.

Note that in \texttt{NeuroSort}, the postsynaptic neurons share the same parameters. When we allow local synaptic modifications to change the connection direction, (binary) radix sort, an algorithm suitable for large-scale data sorting, can be implemented in SNNs. We iterate through the \( b \) bits, applying the \texttt{NeuroSort} to perform bitwise sorting step by step, as outlined in Algorithm \ref{algo:radix}.

\begin{algorithm}
\caption{Neuromorphic (Binary) Radix Sort (\texttt{NeuroRadixSort})}
\label{algo:radix}
\begin{algorithmic}
    \Function{NeuroRadixSort}{$arrIn$}
        \State $\mathrm{arrOut} \gets \mathopen{}\left[  \mathclose{}\right]$
        \State $\mathrm{bitCount} \gets \mathrm{GetMaxBitCount} \mathopen{}\left( \mathrm{arrIn} \mathclose{}\right)$
        \For{$b \in \mathrm{range} \mathopen{}\left( \mathrm{bitCount} \mathclose{}\right)$}
            \State $\mathrm{arrSorted} \gets \mathopen{}\left[  \mathclose{}\right]$
            \State $t \gets 0$
            \While{$\mathrm{len} \mathopen{}\left( \mathrm{arrSorted} \mathclose{}\right) < \mathrm{len} \mathopen{}\left( \mathrm{arrIn} \mathclose{}\right)$}
                \For{$v \in \mathrm{arrIn}$}
                    \If{$v \mathbin{\&} \mathopen{}\left( 1 \ll b \mathclose{}\right) = t$}
                        \State $\mathrm{arrSorted}.\mathrm{append} \mathopen{}\left( v \mathclose{}\right)$
                    \EndIf
                \EndFor
                \State $t \gets t + 1$
            \EndWhile
            \State $\mathrm{arrOut} \gets \mathrm{arrSorted}$
        \EndFor
        \State \Return $\mathrm{arrOut}$
    \EndFunction
\end{algorithmic}
\end{algorithm}

The computational complexity is split into \texttt{GetMaxBitCount} and radix sort. For radix sort, due to the necessity of synaptic modifications, the calculation requires $b \cdot (2+N)$ steps, consuming $b \cdot N$ spikes in total, making it faster but more energy-intensive than \texttt{NeuroSort}. When \( b \) cannot be determined in advance, \texttt{GetMaxBitCount} must use \texttt{NeuroSort} to compute $\max \boldsymbol{x}$, and this introduces a non-negligible cost. Hence, \( b \) is typically pre-set to the minimum bit width needed to represent the data (e.g., 32 bits for \verb|int|).

In particular, when $N \geq 2^b/b$, in scenarios with large data chunks, the overall performance of \texttt{NeuroSort} is superior to that of \texttt{NeuroRadixSort}.
\section{Algorithm Design}
\label{sec:design}

\subsection{Existing Works}

In their work, \cite{Kay2020} introduces a Prim-inspired neuromorphic MST algorithm, achieving a time complexity comparable to the conventional Prim's algorithm \cite{Prim1957}. As detailed in Algorithm \ref{algo:prim-snn}, this method embeds the graph in an SNN with fractional-offset deduplication, \texttt{Deduplicate}, and iteratively identifies the shortest edge connecting a vertex in the MST to a vertex outside the MST, utilizing minimal communication to reconfigure the network between iterations. The \texttt{NeuroSpike} routine requires at most $O\left(\max_{e\in |E|} w_e\right)$ steps per execution, where $|E|$ denotes the edges and $w_e$ represents the weight of the edge $e$. Since the algorithm executes the routine exactly $|V|$ times, it leads to an overall time complexity of $O\left(|V|^2 \cdot \max_{e\in |E|} w_e\right)$ and a space complexity of $O(|V|+|E|)$. By introducing specialized neuromorphic primitives for the MST problem, which achieve asymptotically equivalent resource consumption, \cite{Janssen2024} improves the algorithm to $O(|V|\cdot \max_{e \in |E|} w_e)$. Both implementations share a common energy complexity of $O(|V|^2)$.

\begin{algorithm}
\caption{Neuromorphic Prim's (\texttt{Prim})}
\label{algo:prim-snn}
\begin{algorithmic}
    \Function{NeuroSpike}{$mstVertices, mstEdges, t$}
        \For{$u \in \mathrm{mstVertices}$}
            \Comment{neuromorphic parallelism}
            \For{$\mathopen{}\left( w, \mathrm{\_}, v \mathclose{}\right) \in u.\mathrm{edges}$}
                \If{$w = t$}
                    \State $\mathrm{mstVertices}.\mathrm{add} \mathopen{}\left( v \mathclose{}\right)$
                    \State $\mathrm{mstEdges}.\mathrm{add} \mathopen{}\left( \mathopen{}\left( w, u, v \mathclose{}\right) \mathclose{}\right)$
                    \State \Return $\mathrm{True}$
                        \Comment{stop activity}
                \EndIf
            \EndFor
        \EndFor
        \State \Return $\mathrm{False}$
    \EndFunction
\end{algorithmic}

\begin{algorithmic}
    \Function{NeruoMSTPrim}{$graph$}
        \State $\mathrm{src} \gets \mathrm{RandomChoice} \mathopen{}\left( \mathrm{graph}.\mathrm{vertices} \mathclose{}\right)$
         \State $\mathrm{graph}.\mathrm{edges} \gets \mathrm{Deduplicate} \mathopen{}\left( \mathrm{graph}.\mathrm{edges} \mathclose{}\right)$
        \State $\mathrm{mstVertices} \gets \mathopen{}\left\{ \mathrm{src} \mathclose{}\right\}$
        \State $\mathrm{mstEdges} \gets \mathopen{}\left\{ \mathclose{}\right\}$
        \While{$\mathrm{len} \mathopen{}\left( \mathrm{mstVertices} \mathclose{}\right) < \mathrm{len} \mathopen{}\left( \mathrm{graph}.\mathrm{vertices} \mathclose{}\right)$}
            \State $t \gets 0$
            \While{$\lnot \mathrm{NeuroSpike} \mathopen{}\left( \mathrm{mstVertices}, \mathrm{mstEdges}, t \mathclose{}\right)$}
                \State $t \gets t + 1$
            \EndWhile
        \EndWhile
        \State \Return $\mathrm{mstEdges}$
    \EndFunction
\end{algorithmic}
\end{algorithm}

However, the algorithm restarts neuromorphic activity each time a postsynaptic neuron spikes to ensure that the minimal edge adjacent to the vertices in the MST is consistently identified. The fractional-offset deduplication, proposed together with the algorithm in \cite{Kay2020}, while ensuring that exactly one new neuron spikes during each pass through the while loop after spiking every neuron in \texttt{mstVertices}, also disrupts parallelism. This leads to an expected runtime cost of $\sum_{e \in |E|_\mathrm{MST}} w_e$ steps but does not take advantage of the available inherent parallelism. Additionally, it cannot operate on graphs with multiple edges, as these cannot be hard-coded into SNNs, where each pair of neurons can only be connected with exactly one synapse, according to the complexity framework specifications.

\subsection{Revisiting Kruskal's}
\label{subsec:seq}

Kruskal's algorithm \cite{Kruskal1956} is also a widely used method for finding the minimum spanning tree of a graph. It begins by sorting the edges based on their weights, then iteratively selects the smallest edge and checks if adding it creates a circle. If it does not, the edge is included in the minimum spanning tree, and this process continues until a complete tree is formed. The algorithm can be understood as comprising two main routines: sorting and union-find, which are executed sequentially, as outlined in Algorithm \ref{algo:seq}. A key feature of Kruskal's algorithm is its union-find routine (or disjoint set data structure), which allows for processing of the "find" and "union" operations on edges. However, these operations are susceptible to race conditions \cite{Anderson1991} if multiple edges are handled at the same time, and require thread synchronization or the fallback mechanism\cite{Simsiri2018, Fedorov2023, Jayanti2024}, which is not supported by current neuromorphic primitives. As a result, such operations can only be carried out by suspending neuromorphic activity and dynamically modifying synaptic connections.

\begin{algorithm}
\caption{Neuromorphic Sequential Kruskal's (\texttt{SeqNeuro} or \texttt{SeqRadix})}
\label{algo:seq}
\begin{algorithmic}
    \Function{NeuroUnionFind} {} (
    \\ \hspace{1.5em} {$edges, mstEdges, numMSTEdges$})
        \State $\mathrm{queue}.\mathrm{append} \mathopen{}\left( \mathrm{edges} \mathclose{}\right)$
        \While{$\lnot \mathrm{queue}.\mathrm{empty} \mathopen{}\left( \mathclose{}\right)$}
            \State $\mathopen{}\left( w, u, v \mathclose{}\right) \gets \mathrm{queue}.\mathrm{pop} \mathopen{}\left( \mathclose{}\right)$
            \If{$\mathrm{NeuroFind} \mathopen{}\left( u, v \mathclose{}\right) > 1$}
                \State $\mathrm{NeuroUnion} \mathopen{}\left( u, v \mathclose{}\right)$
                \State $\mathrm{mstEdges}.\mathrm{add} \mathopen{}\left( \mathopen{}\left( w, u, v \mathclose{}\right) \mathclose{}\right)$
                \If{$\mathrm{len} \mathopen{}\left( \mathrm{mstEdges} \mathclose{}\right) = \mathrm{numMSTEdges}$}
                    \State \Return $\mathrm{True}$
                \EndIf
            \EndIf
        \EndWhile
        \State \Return $\mathrm{False}$
    \EndFunction
\end{algorithmic}
\begin{algorithmic}
    \Function{NeuroSeqKruskal}{$graph$}
        \State $\mathrm{edgesSorted} \gets \mathrm{NeuroSort} \mathopen{}\left( \mathrm{graph}.\mathrm{edges} \mathclose{}\right)$
        
        \Comment{or NeuroRadixSort}
        
        \State $\mathrm{mstEdges} \gets \mathopen{}\left[  \mathclose{}\right]$
        \For{$\mathopen{}\left( w, u, v \mathclose{}\right) \in \mathrm{edgesSorted}$}
        \Comment{batch submit}
            \If{$\mathrm{NeuroUnionFind} \mathopen{}\left( \mathopen{}\left[ \mathopen{}\left( w, u, v \mathclose{}\right) \mathclose{}\right] \mathclose{}\right)$}
                \State \Return $\mathrm{mstEdges}$
            \EndIf
        \EndFor
    \EndFunction
\end{algorithmic}
\end{algorithm}

\begin{figure}
    \centering
    \includegraphics[trim=70 560 277 80,clip,width=0.92\columnwidth]{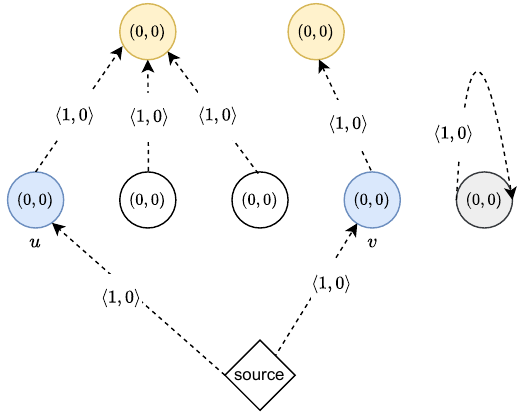}
    \caption{SNN of \texttt{UnionFind} (sequential)}
    \label{fig:unionfind}
\end{figure}

Building on this, we designed a union-find SNN implementation that supports synaptic modifications. The implementation consists of a source with two synapses, $|V|$ neurons, and their respective synaptic connections, as depicted in Figure \ref{fig:unionfind}. It includes a cache queue to store edges temporarily for further processing. Each query retrieves the front element of the queue and modifies the synaptic connections of the source (represented by dashed lines) to point to the two neurons corresponding to the endpoints $u$ and $v$ (i.e., the blue circles). Once the source fires a spike, it propagates through neurons $u$ and $v$, causing them to transmit spikes to their parent neurons (represented by yellow circles). If more than one neuron fires, the activity of all neurons in this SNN is paused, and the synaptic connections are modified according to the principle of union-by-rank and path compression \cite{Tarjan1979,Tarjan1984}. Although synaptic modifications can be performed in parallel, we still define the complexity of a single operation as $\alpha(|V|)$, where $\alpha$ denotes the inverse Ackermann function, refelcting the expected overhead of the overall operation, as suggested in \cite{Fedorov2023}. Initially, the synapses of the neurons point to themselves, and two spikes with distinct timestamps are recorded, as shown by the gray circles in the figure.

\begin{figure}
    \centering
    \includegraphics[trim=60 430 170 85,clip,width=0.92\linewidth]{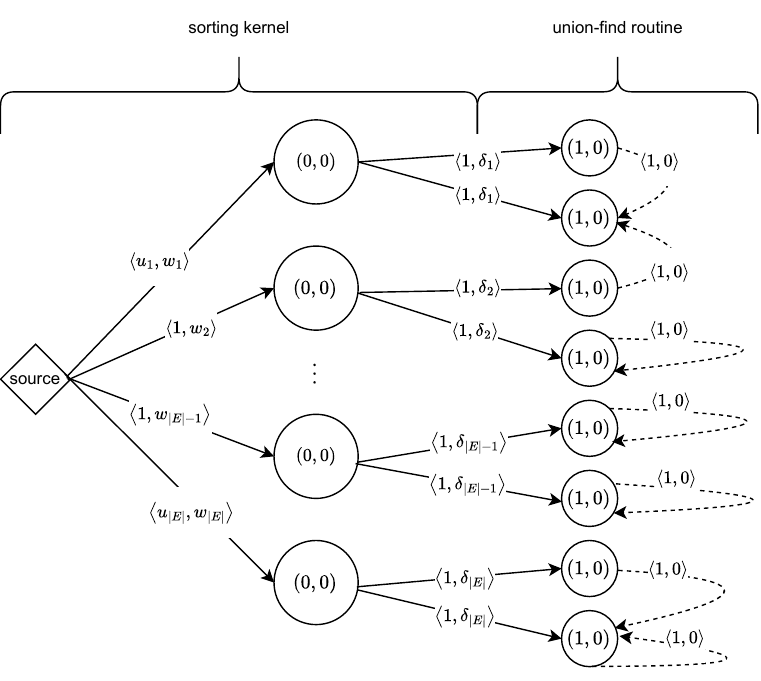}
    \caption{SNN of \texttt{Pipe}}
    \label{fig:pipe}
\end{figure}

The computational cost of the union-find operation is reflected in its execution time, which includes modifying the source synapse for each query, resulting in a cost of $|E|\cdot(2+\alpha(|V|)$. This SNN uses a total of $|V|$ neurons and $|V|+2$ synapses. Each query consumes four spikes (two for $u$ and $v$, and two for their parents), leading to an overall energy cost of $4\cdot|E|$. Under a sequential execution model, the operational cost is the sum of the sorting and union-find costs, as summarized in Table \ref{tab:cost}. In general, for large-scale graphs, sequential approaches using Kruskal's algorithm tend to outperform those using Prim's algorithm.

\subsection{Pipelining In Action}

The performance bottleneck of Kruskal's algorithm arises from the sequential execution of its two fully decoupled stages. This sequential process limits the ability to fully leverage the advantages of event-driven computation. Specifically, in the execution of Kruskal’s algorithm, once the minimum edge is selected, the subsequent union-find query can be triggered through spikes, allowing the sorting kernel to continue executing. In other words, by utilizing \texttt{NeuroSort} for the sorting kernel, we can pipeline Kruskal’s algorithm, thereby enhancing computational efficiency, as illustrated in Algorithm \ref{algo:pipe}.

\begin{algorithm}
\caption{Neuromorphic Pipelined Kruskal's (\texttt{Pipe})}
\label{algo:pipe}
\begin{algorithmic}
    \Function{NeuroPipeKruskal}{$graph$}
        \State $\mathrm{mstEdges} \gets \mathopen{}\left[  \mathclose{}\right]$
        \State $\mathrm{numMSTEdges} \gets \mathrm{len} \mathopen{}\left( \mathrm{graph}.\mathrm{vertices} \mathclose{}\right) - 1$
        \State $t \gets 0$
        \State $\mathrm{done} \gets \mathrm{False}$
        \While{$\lnot \mathrm{done}$}
            \State $\mathrm{edges} \gets \mathopen{}\left[  \mathclose{}\right]$
            \For{$\mathopen{}\left( w, u, v \mathclose{}\right) \in \mathrm{graph}.\mathrm{edges}$}
                
                \Comment{neuromorphic parallelism}
                \If{$w = t$}
                \Comment{batch collect}
                    \State $\mathrm{edges}.\mathrm{append} \mathopen{}\left( \mathopen{}\left( w, u, v \mathclose{}\right) \mathclose{}\right)$
                \EndIf
            \EndFor
            \State $\mathrm{done}$ $\gets$ $\mathrm{NeuroUnionFind} \mathopen{}$(\newline
        \hspace*{5em} $\mathrm{edges}, \mathrm{mstEdges}, \mathrm{numMSTEdges} \mathclose{}$)
            
            \Comment{batch submit}
            \State $t \gets t + 1$
        \EndWhile
        \State \Return $\mathrm{mstEdges}$
    \EndFunction
\end{algorithmic}
\end{algorithm}

The SNN structure of the pipelined Kruskal's algorithm is shown in Figure \ref{fig:pipe}, consisting of two main components: the sorting kernel and the union-find routine. In this structure, each group of neurons and synapses in the sorting kernel corresponds to an edge in the graph, with each neuron connected to the corresponding endpoint neuron in the union-find routine via a fixed synapse (referred to as a "pipe"). We define a time step as "valid" if spikes are generated by the sorting kernel during that time step. Suppose that at the $j$-th valid time step, $s_j$ edges are traversed. The neurons corresponding to these edges will generate spikes, and before these are propagated, we modify the delay of the corresponding "pipes" to be incremented. After modifying a pipe, the spike is immediately transmitted, ensuring that the spikes are submitted sequentially.

To better evaluate the computational overhead of the entire pipeline, we focus on the startup and completion time overhead of the union-find routine. For the \( j \)-th valid time step \( t_j \), the union-find routine needs to wait \( \Delta t_j = t_j - t_{j-1} \) steps before it can start. If \( s_j \) edges need to be modified, it takes \( 2 \cdot s_j \) steps to configure the delay for each synapse. After each pair of synapses is configured, the spike is immediately submitted. Since neuromorphic activity needs to be paused, each submission requires \( \alpha(|V|) \) time steps to complete. Therefore, the computational overhead for the current time step is \( \Delta t_j + 2 \cdot s_j + s_j \cdot \alpha(|V|) \). The total overhead for the entire pipeline is the sum of the overheads for all valid time steps. The sorting kernel uses \( |E| \) neurons and synapses, while the union-find routine uses \( |V| \) neurons and synapses. The "pipes" use \( 2 \cdot |E| \) synaptic connections. During execution, the neurons in the sorting kernel send spikes to the subsequent two neurons to trigger the union-find routine, consuming a total of \( 2 \cdot |E| + 4 \cdot |E| \) spikes. These overheads are also summarized in Table \ref{tab:cost}.

\begin{table*}[ht]
    \centering
    \begin{tabular}{lrrrr}
    \toprule
    Approaches & Time (steps) & Neuron Count & Synapse Count & Spike Count \\
    \midrule
    Prim & $\sum_{e\in|E|_\mathrm{MST}} w_e$ & $|V|$ & $|E|$ & $|V|^2$ \\
    \texttt{SeqNeuro} & $\max_{e\in|E|} w_e + |E|\cdot\left(2+\alpha(|V|)\right)$ & $|E|$ + $|V|$ & $|E| + (2 + |V|)$ & $|E| + 4\cdot |E|$ \\
    \texttt{SeqRadix} & $b\cdot (2+|E|)+|E|\cdot\left(2+\alpha(|V|)\right)$ & $|E|+|V|$ & $|E| + (2 + |V|)$ & $b\cdot |E| + 4\cdot |E|$ \\
    \texttt{Pipe} & $\sum_{s_j > 0} \left(\Delta t_j + 2\cdot s_j + s_j\cdot\alpha(|V|)\right)$ & $|E|+|V|$ & $|E|+2\cdot |E|+|V|$ & $2\cdot|E|+4\cdot |E|$ \\
    \bottomrule
    \end{tabular}
    \caption{A summary of execution overheads for different MST approaches}
    \label{tab:cost}
\end{table*}

As shown in Table \ref{tab:cost}, compared to Prim's algorithm, Kruskal's approaches accelerate execution by utilizing more neurons and synapses, trading off resource usage for higher performance. However, when $|E| < |V|^2$, Kruskal’s overall power consumption is lower than of Prim's. Notably, due to the use of "pipes", although the actual neuron firing count is the same as in \texttt{SeqNeuro}, there are an additional $|E|$ spikes fired in \texttt{Pipe}, but still fewer than in \texttt{SeqRadix}. Pipelining typically achieves significant performance improvements in most scenarios. However, when the time required to find the maximum edge of the MST is greater than or equal to the time needed to complete the sorting itself, the pipelining effect becomes less effective than sequential Kruskal’s, resulting in a bottleneck in the entire pipeline. We will discuss this phenomenon in more detail in Section \ref{sec:results}, supported by comprehensive experiments.
\section{Experiments}
\label{sec:results}

\subsection{Environment Setup}

We employ the PyTorch library \footnote{https://github.com/pytorch/pytorch} and the SpikingJelly framework\footnote{https://github.com/fangwei123456/spikingjelly}, along with various third-party packages, to implement the neuromorphic kernels described in this study. The experiments are carried out in the environment outlined in Table \ref{table:setup} and are executed on a GPGPU to facilitate faster simulations. We carry out thorough sanity checks to ensure the accuracy of the kernels in comparison to industry standards such as NetworkX \footnote{https://github.com/networkx/networkx}.

\begin{table}[!ht]
    \centering
    \begin{tabular}{|l|p{0.8\linewidth}|} 
        \hline
        \textbf{CPU} & Intel(R) Xeon(R) Platinum 8358P \\ \hline
        \textbf{GPU} & NVIDIA GeForce RTX 4090 (driver v550.120) \\ \hline
        \textbf{OS} & Ubuntu 24.04.1 \\ \hline
        \textbf{Python} & Python 3.12.8 + conda 25.1.1 (miniconda)\\ \hline
        \textbf{Packages} & PyTorch 2.6.0 with CUDA 12.4 + SpikingJelly 0.0.0.0.14 + Scipy 1.15.2 + NetworkX 3.4.2 + nx-cugraph 24.12  \\ \hline
    \end{tabular}
    \caption{Environment configuration}
    \label{table:setup}
\end{table}

\subsection{Performance on DIMACS10 matrices}

We evaluate different approaches to Kruskal's algorithm on the large-scale graphs from the DIMACS10 dataset\footnote{https://sparse.tamu.edu/DIMACS10} \cite{SparseSuite}, comparing them against the state-of-the-art neuromorphic implementations based on Prim's algorithm \cite{Kay2020,Janssen2024}. Our analysis consists of 20 undirected weighted graphs from the DIMACS10 dataset, each containing close to or greater than 1 million nonzero elements. This selection aims to replicate data-intensive computational scenarios. The characteristics of these graphs are summarized in Table \ref{tab:dimacs}. This setup enables us to assess the performance of various neuromorphic approaches in the search for the MST of large-scale graphs. 

\begin{figure}
    \centering
    \includegraphics[trim=12 12 10 15,clip,width=0.92\linewidth]{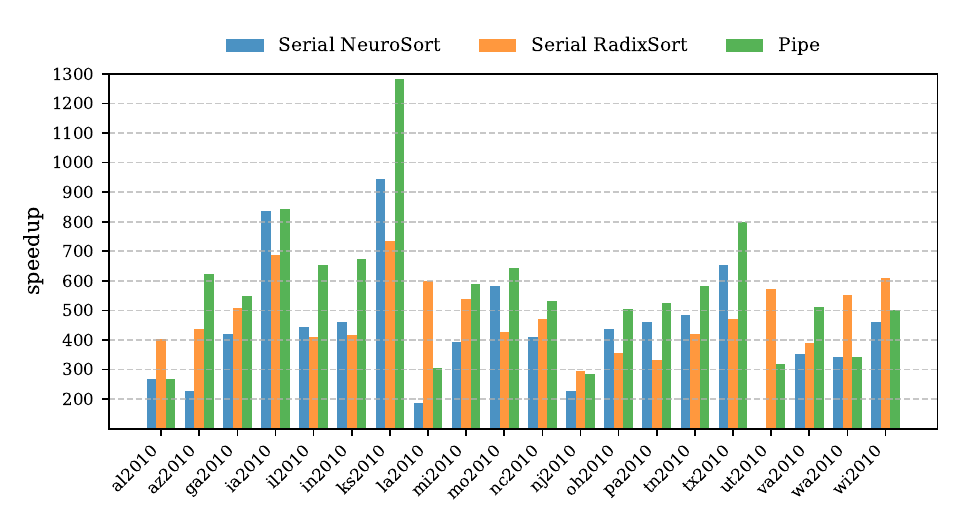}
    \caption{Speedup comparison between \texttt{SeqNeuro}, \texttt{SeqRadix} and \texttt{Pipe} on DIMACS10 dataset}
    \label{fig:speedup}
\end{figure}

Figure \ref{fig:speedup} presents a comparison of the speedup achieved by \texttt{SeqNeuro}, \texttt{SeqRadix}, and \texttt{Pipe} relative to the state-of-the-art Prim's implementations. Notably, \texttt{Pipe} outperforms \texttt{SeqNeuro} and \texttt{SeqRadix} in 14 out of 20 graphs tested. Further analysis aims to identify the bottlenecks in \texttt{Pipe} for the remaining 6 graphs: \verb|al2010|, \verb|la2010|, \verb|nj2010|, \verb|ut2010|, \verb|wa2010|, and \verb|wi2010|.

\texttt{Pipe} functions by triggering subsequent union-find queries while traversing the MST edges through neuromorphic sorting. If the time required to enumerate all MST edges exceeds the time needed for sorting, this process can become a bottleneck for the entire pipeline. To validate this, we compared the time taken for radix sorting on all edges with the time taken to enumerate all MST edges.

\begin{figure}
    \centering
    \includegraphics[trim=50 12 50 15,clip,width=0.92\linewidth]{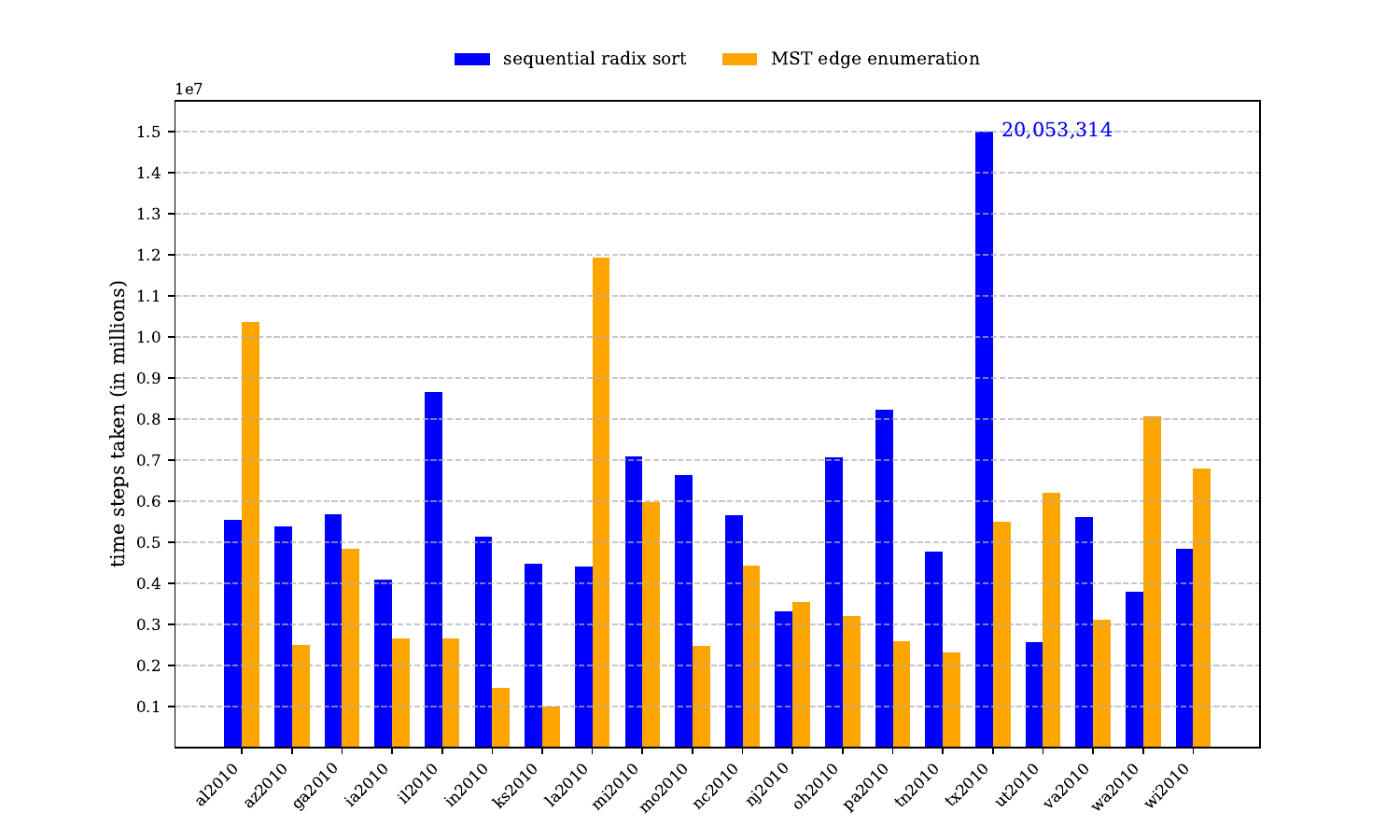}
    \caption{Time taken comparison between radix sort and MST edge enumeration on DIMACS10 dataset}
    \label{fig:anlz}
\end{figure}

The result, shown in Figure \ref{fig:anlz}, indicate that for these six graphs, identifying the maximum edge of the MST takes significantly longer than the radix sort itself. This suggests that in \texttt{Pipe}, the enumeration of the MST's maximum edge using the \texttt{NeuroSort} paradigm impairs the pipeline's efficiency, leading to lower efficiency compared to \texttt{SeqRadix}. In other graphs where radix sort takes longer, \texttt{Pipe} achieves a speedup ratio ranging from 1.084x to 1.75x, with a median ratio of 1.421x when compared to \texttt{SeqRadix}.

\begin{table}[ht]
    \centering
    \begin{tabular}{lrrl}
    \toprule
    Graph & Number of vertices & Number of edges & Edge weight distribution \\
    \midrule
    al2010 & 252266 & 615241 & [9, 10370522] \\
    az2010 & 241666 & 598047 & [9, 14067507] \\
    ga2010 & 291086 & 709028 & [9, 7859220] \\
    ia2010 & 216007 & 510585 & [9, 2716349] \\
    il2010 & 451554 & 1082232 & [9, 7498690] \\
    in2010 & 267071 & 640858 & [9, 4176383] \\
    ks2010 & 238600 & 560899 & [29, 2609740] \\
    la2010 & 204447 & 490317 & [14, 21666664] \\
    mi2010 & 329885 & 789045 & [9, 11678592] \\
    mo2010 & 343565 & 828284 & [9, 3332235] \\
    nc2010 & 288987 & 708310 & [9, 7190375] \\
    nj2010 & 169588 & 414956 & [9, 5164756] \\
    oh2010 & 365344 & 884120 & [9, 4639190] \\
    pa2010 & 421545 & 1029231 & [9, 3946650] \\
    tn2010 & 240116 & 596983 & [9, 3605330] \\
    tx2010 & 914231 & 2228136 & [14, 10149954] \\
    ut2010 & 115406 & 286033 & [41, 23553227] \\
    va2010 & 285762 & 701064 & [10, 6753024] \\
    wa2010 & 195574 & 473716 & [9, 8072023] \\
    wi2010 & 253096 & 604702 & [22, 7805919] \\
    \bottomrule
    \end{tabular}
    \caption{Characteristics of large-scale graphs in DIMACS10 dataset}
    \label{tab:dimacs}
\end{table}

The analysis sheds light on the decision between using \texttt{Pipe} and \texttt{SeqRadix} by estimating the core speedup comparison. Specifically, \texttt{Pipe} is better suited for low-power scenarios with a significantly higher density of small edge weights. In particular, if the maximum weight of the MST cannot be determined in advance, opting for \texttt{Pipe} is always the preferable choice.
\section{Discussion}
\label{sec:discuss}

We primarily discuss the feasibility of implementing structural plasticity on existing neuromorphic platforms. Structural plasticity requires the SNNs to exhibit self-generative properties, meaning that synaptic connections are adjusted based on factors such as neuron firing rates and spike event density \cite{vanOoyen2017}, optimizing overall network performance. Unlike traditional artificial neural networks (ANNs) or conventional SNNs that disconnect synaptic connections by setting specific synaptic weights to zero, this type of SNN necessitates the dynamic allocation and recycling of synaptic connections. For platforms that do not support dynamic synaptic resource management, such as memristor crossbars \cite{Mustafazade2024}, implementing a union-find SNN, as proposed in Section \ref{subsec:seq}, could require resources on the order of $|V|^2$, significantly exceeding the resources needed to directly embed a graph structure using Prim’s algorithm, despite both sharing a common space complexity of $O(|E|+|V|)$.

Fortunately, several neuromorphic platforms currently support this type of learning rule. \cite{George2017} proposed a synaptic resource allocation and recycling algorithm, which was implemented on an FPGA as a co-processor to assist the neuromorphic chip ROLLS. \cite{Bogdan2018} implemented a structural plasticity framework on the SpiNNaker platform, demonstrating improvements in tasks such as topographic map generation through synaptic rewiring and the Spike-Timing-Dependent Plasticity (STDP). \cite{Billaudelle2021} implemented algorithms for synaptic pruning, reassignment, and correlation-driven weight updates on the BrainScaleS-2 platform, performing supervised learning on a digital processor to demonstrate its ability to optimize network topology. Additionally, an increasing number of generic neuromorphic simulation platforms \cite{Fang2018,Lee2021,Chen2022,Liu2024} now enable the realization of synaptic rewiring through customized operator operations, facilitating more efficient design exploration for neuromorphic algorithms.

In proposing these designs, we carefully consider the overhead introduced by structural plasticity and incorporate it into the computational complexity analysis of our algorithm. Experimental results further demonstrate that, despite the additional overhead associated with these operations, our pipelined Kruskal's algorithm still outperforms the Prim-based implementations.
\section{Conclusion}

Building on prior work analyzing data movement in neuromorphic systems, we propose a revision to the existing neuromorphic computational complexity model, accounting for the overhead introduced by the dynamic synaptic plasticity during runtime. Leveraging these primitives, we have designed a neuromorphic union-find routine based on the SNNs. 

During the design phase, we identified a key bottleneck in Kruskal’s algorithm: the sequential execution of its two fully decoupled stages, which prevents the efficient exploitation of the event-driven computation inherent in neuromorphic systems. To address this limitation, we propose pipelining Kruskal’s algorithm using spike-driven neuromorphic sorting. This novel design is difficult to implement within conventional computing architectures, underscoring the potential advantages of neuromorphic computing. In our approach, each time the sorting kernel selects the minimum-weight edge, it is immediately submitted to the union-find routine for processing while the sorting kernel continues its execution in parallel.

We analyze the computational complexity of three different approaches and evaluate their performance on the DIMACS10 dataset alongside Prim’s algorithm. Our results indicate that the pipelined approach achieved speedups ranging from 269.67x to 1283.80x, with a median of 540.76x, surpassing the sequential approaches in most cases. If the time required to enumerate the MST edges is shorter than the time needed for a full sorting of edge weights, the pipelined approach avoids bottlenecks.

\bibliographystyle{ieeetr}
\bibliography{main}

\end{document}